\documentstyle[latex8]{article}
\begin{document}
\textwidth 17.5cm
\textheight 22.54cm

\title{Polynomial Simulations of Decohered Quantum Computers}

\author{D.~Aharonov\thanks{
Institutes of Physics and Computer science, The Hebrew University,
Jerusalem, Israel, E-mail: doria@cs.huji.ac.il}
 \and M.~Ben-Or\thanks{
Institute of Computer science,
The Hebrew University, Jerusalem,
Israel, E-mail: benor@cs.huji.ac.il}}
\date{}

\maketitle
\thispagestyle{empty}
\begin{abstract}
Recently it has become clear, that a key issue in 
 quantum computation is understanding 
how interaction with the environment, or ``decoherence'',
 effects the computational power of quantum computers.
We adopt the  standard physical
method of describing systems which are
interwound with their environment by ``density matrices'',
and within this framework define a model of decoherence in quantum 
computation.

Our results show that the computational power of decohered quantum computers
depends strongly on the amount of parallelism in the computation.
We first present a simulation of decohered sequential quantum 
computers, on a 
classical probabilistic Turing machine, and prove that the expected slowdown of
this simulation is polynomial in time and space of the quantum computation,
for any non zero decoherence rate. Similar results hold for 
Quantum computers that are allowed to operate on logarithmic 
number of qubits at a time.

For decohered quantum circuits
(with local gates), the situation is more 
subtle and depends on the decoherence rate, $\eta$.
We find that our simulation  is efficient for circuits
with decoherence rate $\eta$ higher than some constant $\eta_{1}$,
but exponential 
for a general (random) circuit subjected to decoherence rate 
lower than some constant $\eta_2$.
The transition from exponential cost to polynomial cost
 happens in a short range of decoherence rates.
We use computer experiments to exhibit the
phase transitions in various quantum circuits.
 \end{abstract}

\newtheorem{theo}{Theorem}
\newtheorem{lemm}{lemma}
\newtheorem{conj}{conjecture}
\newtheorem{deff}{definition}
\newcommand {\wbox} {\mbox {$\sqcap$\llap {$\sqcup$}}}
\newcommand {\bbox} {\vrule height7pt width4pt depth1pt}

\newcommand {\zo} [1] {\{0,1\}^{#1}}

\section{Introduction}

Quantum Turing Machines\cite{Deu85,BV93} 
and Quantum Circuits\cite{Deu89,Yao93} challenge the so called 
``polynomial Church thesis'' which asserts that 
 ``randomized Turing machines
can simulate with polynomial slowdown any computational device.''
In particular Shor's quantum factoring algorithm\cite{Shor94} provides
 within this theoretical
 framework, an efficient solution,
 to a problem for which no classical polynomial algorithm is known.

It is yet unclear whether and how quantum computers will
 be physically realizable,\cite{Loyd93,DiV95}
but as any physical system, they {\em in principle}
will not  be ideally 
isolated, and will to some extant interact with their
environment. Such interaction causes the state of the
 computer to be interwound, or entangled,   
with the state of the environment, a process called ``decoherence''
\cite{Zur91,Unr94,PSE95}.
A real challenge to the ``polynomial church thesis'' would come
from {\em decohered} quantum computers.
 Early works\cite{Chu95}
 showed that the effects of {\em decoherence}
on the quantum computation can not be ignored, and that decoherence
 may limit the applicability of
quantum algorithms.
Thus, understanding the computational
power of decohered quantum computers is essential.


Decoherence  takes us out  of  the standard model  of
quantum  computers  described  by  {\it pure  states},
  as  the  state of  a decohered  computer
is  in  general  a {\it  mixed  state}.  
Mixed states are used when there is only partial knowledge about the system,
due to the fact that the system is interwound with it's environment. 
Such a state can be represented as a 
 probability $p_{k}$ for the system to be in the 
 pure state $|\alpha_{k}\rangle$. 
This description  is  not unique  
 and  instead we  use a unique representation  
 of mixed states called {\em density matrices}\cite{Saq},
to describe the state of the quantum computer\cite{AN96}.
%
%

 In order to incorporate decoherence into the model, we 
need to add a clock  to the circuit
since the process of errors is dynamic and depends on time.
A circuit is associated a  {\it timing} which   indicates the time  step for
each gate  to operate.
To add decoherence, the following assumptions are made: 
(1)  Each qubit decoheres independently  (single qubit faults), 
and  (2) No decoherence 
takes place inside the gates.
Each qubit  decoheres,
 or undergoes  a fault with probability $\eta$ per step, and $\eta$
is referred to as the {\it  decoherence  rate}.
The triplet of a quantum circuit, a timing of the circuit,
 and a decoherence rate construct together what we call a 
{\it quantum medium}, which  
operates subjected to different types of
single qubit faults.
The list of times and places where faults had occured, namely
 the {\it fault path},
 is random, and 
naturally, the function that the medium computes is the weighted average
over the outputs for each possible {\it fault path}.
This model for decoherence is richer than it seems at first glance:
it is equivalent to noise caused by weak interactions with the environment,
and, 
 to a model in which there are many possible kinds of faults
 that can occur.
Our results therefore apply to these kinds of decoherences too.

In this paper we exhibit
upper bounds on the computational power of quantum
 mediums subjected to faults. 
As a first observation, note  that regarding upper bounds,
 it is 
enough to restrict the discussion to faults of a specific type,
which we choose to be {\it collapses}
of a single qubit, or in other words: measurements
of a single qubit in some basis.
It suffices to deal with collapses because of two reasons:
 The first is that 
if a quantum medium claims
to compute a function fault tolerantly, it is supposed 
to compute it when subjected to any fault.
The second, and maybe more significant reason, is that
   it is reasonable
to assume that every physical realization of quantum computers will
be subjected to some extent to collapses\cite{Unr94}.

We present a simulation of a quantum medium subjected 
to single qubit collapses, on a classical probabilistic Turing machine.
The simulation chooses, with the appropriate probability,
a fault path. Then it keeps  track of the development in time 
of the density matrix of the medium
according to this path, and gets a final density matrix.
The simulation imitates measurements on this density matrix
and outputs the result.
Since the algorithm chooses the path with the correct weight,
the distribution over the results will be as that of the
medium.
How does the simulation keep track on 
the development of the density matrix, which is exponential in size?
Trying to be efficient, the simulation presents this matrix as 
a tensor product of smaller density matrices. 
The input string is described as a tensor product 
of the density matrices
of each qubit,
but if no faults occured,  
 the computer can soon entangle all the qubits together, 
with no way to view their exponential density matrix 
 as a tensor product of smaller matrices.
Collapses prevent this from happening, since
 after a qubit collapses to one of it's basic states
 it is no longer  entangled to the
other qubits,
and can be described by it's own density matrix.

We are interested in the expected cost of the simulation.
The results emphasize
the significance of  the amount of parallelism in the computation.
 To gain some intuition,   
one can view the decohered quantum computation as a struggle between
two forces: The computation, which
 tries to use the exponential 
dimension of the Hilbert space, by entangling as many qubits as it can,
and the decoherence,
which destroys this entanglement by collapses. 
>From this point of view, a sequential computer obviously
 can not win the battle - 
each time step it applies  only one local operation,
 but $\theta(n)$  qubits collapse.
The picture in mediums with general parallelism is different,
since both forces that compete have linear power:
$\theta(n)$ gates can be applied each time step,
and the decoherence collapses a percentage of the n qubits.
Since the competition here is even, the question which force wins in this
struggle is delicate and requires careful consideration.

Our results show that:

$\bullet$ The expected slowdown of our simulation applied
on 
{\bf mediums with O(log(n)) parallelism}, with any non zero decoherence rate,
 is polynomial in the number of qubits and time steps.  
This holds
also for multi head Quantum Turing machines that are allowed to operate 
 on $O(log(n))$ qubits each time step.
 
The mathematical techniques involved in the case of 
{\bf mediums with general parallelism} are more  sophisticated.
In our analysis we
  assume that the gates are of fan-in not bigger than two\cite{DiV2},
but the results can be extended to any constant fan-in:

$\bullet$ {\bf High decoherence rate:} The simulation of mediums with
decoherence rate higher than some constant is efficient.
Single-qubit-gates can not
entangle different qubits, and therefore do not participate in the 
``struggle'' against decoherence.
Hence if two-qubits-gates are applied ``rarely'',
the cost is polynomial already at lower decoherence rates.

$\bullet$ {\bf Low decoherence rates:} The cost of simulating
 a random circuit is exponential, for 
decoherence rates  lower than some constant.

$\bullet$  {\bf Phase transition in the cost:}
Results regarding phase transitions\cite{fri} imply
that the cost of simulating a random circuit transforms 
from exponential to polynomial in a very short range 
of decoherence rates.
As for other  quantum circuits,
if the cost transforms from exponential to polynomial
this transition is sharp~(by extension 
of \cite{fri}, private communication).

\smallskip
We used computer experiments to show that 
in the case of a random circuit, the phase transition  
occurs  at $\eta_{0} \approx 0.63$.
We also looked at 
 a medium which is a one dimensional array of qubits, 
where  gates are applied 
only on nearest neighbors, and found
that in this case the transition occurs already at $\eta \approx 0.50$.

{\bf Organization of paper:} In section 2 we recall the basic definitions
of quantum circuits and mixed states. In section 3 we define the
model of quantum mediums with faults.
Section 4 is devoted to the simple case of simulating sequential mediums.
In  section 5 we concentrate on simulating mediums with
general parallelism.
Section 6 discusses consequences of this work.

\section{Quantum circuits with mixed states}
In this section we
 recall some physical definitions of quantum systems
of n two-state particles, Hilbert space, computational basis, pure states,
 mixed states and density matrices\cite{Saq}.
We then describe the model of quantum circuits with mixed states,
defined in \cite{AN96}.

\subsection{\bf Pure states and mixed states} 
\noindent
{\bf Pure states:}
 We deal with systems of  n two-state quantum
particles, or ``qubits''. The {\em pure state} of such a system 
is a unit vector, denoted $|\alpha\rangle$,
in the Hilbert space\footnote{A Hilbert space is
 a vector space with 
an inner product} $\cal{C}$$^{2^{n}}$
i.e. a $2^{n}$ dimensional complex space.
We view  $\cal{C}$$^{2^{n}}$ as a
 tensor product n two dimensional spaces, each corresponding to a qubit:
$\cal{C}$$^{2^{n}}= \cal{C}$$^{2}\otimes...\otimes\cal{C}$$^{2}$.
As a basis for   $\cal{C}$$^{{2}^{n}}$,
we use the $2^{n}$ orthogonal {\it basic states}:
 $|i\rangle=|i_{1}\rangle\otimes
|i_{2}\rangle....\otimes|i_{n}\rangle,0\le i< 2^{n}$,
 where $i$ is in binary representation.
and each $i_{j}$ gets 0 or 1.
Such a state corresponds to 
the j'th qubit being in the state 
   $|i_{j}\rangle$.
A pure state  $|\alpha\rangle\in\cal{C}$$^{{2}^{n}}$ 
is a {\em superposition}
 of the basic states:  
$|\alpha\rangle = \sum_{i=1}^{2^{n}} c_{i}|i\rangle$, 
 with $\sum_{i=1}^{2^{n}} |c_{i}|^{2}=1$. $|\alpha\rangle$ corresponds
to the vector 
 $v_{\alpha}=(c_{1},c_{2},...,c_{2^{n}})$.
 $v_{\alpha}^{\dagger}$, the complex conjugate of  $v_{\alpha}$,
is denoted  $\langle\alpha|$.
The inner product between $|\alpha\rangle$ and $|\beta\rangle$
is $\langle\alpha|\beta\rangle=
 (v_{\alpha},v^{\dagger}_{\beta})$.
The matrix  $v_{\alpha}^{\dagger}v_{\beta}$
 is denoted as  $|\alpha\rangle\langle\beta|$.
An isolated system of n qubits 
develops in time by a unitary matrix\footnote{Unitary
 matrices  preserve the
norm of any vector and satisfy the condition 
$U^{-1}=U^{\dagger}$}
 of size $2^{n} \times 2^{n}$:
 \( |\alpha(t_{2})\rangle = U|\alpha(t_{1})\rangle.\)
A quantum system in $\cal{C}$$^{{2}^{n}}$ can be {\em observed} by 
{\em measuring} the system.
An  important
 measurement is a {\it basic 
measurement} of
 a qubit $q$, of which the possible outcomes are $0,1$.
For the state  $|\alpha\rangle=\sum_{i=1}^{2^{n}} c_{i}|i\rangle$,
the  probability for outcome $0$ is $p_{0}= \sum_{i, i|_{q}=0}|c_{i}|^{2} $
and the state of the system 
will {\em collapse} to 
$|\beta\rangle=\frac{1}{p_{0}}\sum_{i, i|_{q}=0} c_{i}|i\rangle$,
 (the same for $1$).
In general,
an {\em observable} $O$ over $\cal{C}$$^{{2}^{n}}$  is  an
 hermitian\footnote{ An hermitian matrix $H$ satisfies $H=H^{\dagger}$} matrix, of size $2^{n}\times 2^{n}$. 
To apply a measurement of $ O$  on 
a pure state $|\alpha\rangle \in \cal{C}$$^{{2}^{n}}$.
write $|\alpha\rangle $  uniquely 
as a superposition
 of unit eigenvectors of $O$:
\(|\alpha\rangle= \sum_{i} c_{i}|o_{i}\rangle\),
where $|o_{i}\rangle$ have different eigenvalues. 
 With probability $ | c_{i}|^{2}$  the  measurement's outcome
 will be the eigenvalue of $|o_{i}\rangle$,
and the state  will {\it collapse} to $|o_{i}\rangle $.
A unitary operation $U$ on $k$ qubits  
 can be applied  on n qubits,
$n\geq k$, by taking the extension $\tilde{U}$  of $U$,
i.e. the tensor product of $U$ with  an identity matrix on
 the other qubits.  The same applies for an observable $O$ to give
$\tilde{O}$.
%

\noindent
{\bf Mixed states:}
 A system which is not ideally isolated  from 
it's environment is described by a {\em mixed state}.
There are two equivalent description of mixed states:
mixtures and density matrices.
Two different density matrix 
can be distinguished by a measurement,
and two systems with the same density matrix are quantumly 
indistinguishable.
In contrast, the mixture description is not unique:
 different mixtures may yield the same
 density matrix.
We use density matrices in this paper.

\noindent
{\bf mixtures:}
A system in the {\bf mixture}  $\{\alpha\}=\{p_{k},|\alpha_{k}\rangle\}$
is  with probability $p_{k}$
in the pure state $|\alpha_{k}\rangle$.
The rules of development in time and measurements 
for mixtures are 
obtained by applying {\bf classical} probability to
the rules 
for pure states.
A unitary matrix $U$ transforms a mixture $\{p_{k},|\alpha_{k}\rangle\}$
 to 
 $\{p_{k},U|\alpha_{k}\rangle\}$.
 To apply an observable O on the mixture 
 $\{\alpha\}=\{p_{j},|\alpha_{j}\rangle\}$.
write each pure state  as a sum of eigenvectors of O with
{\it different} eigenvalues:
\(|\alpha_{j}\rangle=\sum_{i} c_{j}^{i}|o_{j}^{i}\rangle\).
The probability to get an eigenvalue $\lambda_{i}$ is 
\(Pr(\lambda_{i})=\sum_{j} p_{j}  |c_{j}^{i}|^{2}\). In 
the resulting mixture, under the condition the result was  $\lambda_{i}$,
 each pure state has collapsed  to 
it's part that is consistent with $\lambda_{i}$, 
and the conditioned probabilities are computed classically:
 \(\{\alpha\}_{\lambda_{i}}=
\{\frac{|c_{j}^{i}|^{2}p_{j}}{\sum_{j}|c_{j}^{i}|^{2}
 p_{j}},|o_{j}^{i}\rangle\}.\)
One can also apply an unconditioned  measurement, i.e. 
 not condition on the outcome of the measurement but instead 
 average on the possible outcomes.
This gives the mixed state $O\circ \{\alpha\}=\{Pr(\lambda_{i}),\{\alpha\}_{\lambda_{i}},
\}$.

\noindent
{\bf Density matrices:}
A density matrix $\rho$ 
on $\cal{C}$$^{2^{n}}$ is an hermitian positive semi definite complex matrix
 of dimentions $2^{n}\times 2^{n}$,
with $tr(\rho)=1$.
A pure state $|\alpha\rangle=\sum_{i} c_{i}|i\rangle$ 
is associated the density matrix
 \(\rho_{|\alpha\rangle} = |\alpha\rangle\langle\alpha|\) i.e.  
\(\rho_{|\alpha\rangle}(i,j)= c_{i}c_{j}^{*}.\)
A mixture 
$\{\alpha\}=\{p_{l},|\alpha_{l}\rangle\}$,
is  associated the density matrix :
\(\rho_{\{\alpha\}} = \sum_{l} p_{l} \rho_{|\alpha_{l}\rangle}.\)
The operations on a density
matrix are defined such that  the correspondence to mixtures is preserved.
If a unitary matrix $U$ transforms the mixture 
\(\{\alpha\}=\{p_{l},|\alpha_{l}\rangle\}\) to
\(\{\beta\}=\{p_{l},U|\alpha_{l}\rangle\}.\)
Then
 \(\rho_{\{\beta\}} = \sum_{l}  p_{l}
 U|\alpha_{l}\rangle\langle\alpha_{l}|U^{\dagger}=
U\rho_{\{\alpha\}}U^{\dagger}.\)
Applying a measurement of an observable O
on $\rho$, which is written in a basis of eigenvectors $v_{i}$ of $O$.
gives, the probability for an outcome $\lambda$ is
 the sum of the diagonal terms of $\rho$, which relate to 
the eigenvalue $ \lambda$: 
$pr(\lambda)=\sum_{i=1}^{2^{n}} \rho_{v_{i},v_{i}} \delta(\lambda_{i}
=\lambda)$.
conditioned that the outcome is the eigenvalue $\lambda$,
the resulting density matrix is $O_{\lambda}\circ(\rho)$, which we get
by  first
putting to zero all rows and columns in $\rho$, which relate 
to eigenvalues different from
$\lambda$, and then renormalizing this matrix to trace one. 
Without conditioning  on the outcome 
the resulting density matrix will be  
\(O\circ(\rho)=\sum_{k}
 Pr(\lambda_{k}) O_{\lambda_{k}}\circ(\rho). \)
which differs from  $\rho$, only in
 that the entries in  $\rho$ which connected between
 different eigenvalues are put to zero. 
Given a density matrix $\rho$ of n qubits,
 the reduced density matrix of a subsystem,$ A$,
 of, say, $m$ qubits is defined as an average over the states of
the other qubits:   
 \( \rho|_{A}(i,j)= \sum_{k=1}^{2^{n-m}} \rho(ik,jk)\).

\subsection{\bf Quantum circuits with mixed states}
We describe the model of quantum circuits\cite{Deu89,Yao93},
 with mixed states\cite{AN96}:
{\em A quantum unitary gate} of order $k$ is a complex unitary 
matrix of size $2^{k} \times 2^{k}$. 
A density matrix $\rho$ will transform
by the gate to  \(g\circ\rho = \tilde{U}\rho\tilde{U}^{\dagger}\),
where $\tilde{U}$ is the extension of $U$.
Using density matrices one can also define a non unitary gate:
{\em A measurement gate} of order $k$ is a complex hermitian 
matrix of size $2^{k} \times 2^{k}$.
A density matrix $\rho$ will transform
by the gate to  $g\circ\rho = \tilde{O}\circ(\rho)$. 
{\em A Quantum circuit} 
 is a directed acyclic graph
with n inputs and n outputs. 
 Each node $v$ in the graph is labeled by a quantum gate $g_{v}$.
The in-degree and out-degree of $v$ are equal to the order of $g_{v}$.
Some of the outputs are labeled ``result'' to indicate that
these are the qubits that will give the output of the circuit.
The wires in the circuit correspond to qubits.
An initial density matrix $\rho$ 
transforms by a circuit $Q$ to a 
 final density matrix  $Q\circ \rho =
g_{t}\circ...\circ g_{2}\circ g_{1}\circ \rho$,
where the gates  $ g_{t}...g_{1}$ are applied in a topological order. 
For an input string $i$,
the initial density matrix is $\rho_{|i\rangle}$.
The output of the circuit is the outcome of applying basic measurements
of the result qubits, on the final density matrix  
$Q\circ\rho_{|i\rangle}$. Since the outcomes of measurements
 are random, the function that the circuit computes is a 
{\it probabilistic function}, i.e. for input $i$ it outputs 
strings according to a distribution which depends on $i$.


\section{Quantum mediums with faults}
In this section we define quantum mediums subjected
to  probabilistic single qubit faults of one type.
We then show that the probabilistic fault model
is equivalent to a deterministic weak fault model and to a model
with many possible types of faults.

\subsection{\bf Quantum mediums with probabilistic faults}
In order to incorporate faults into quantum circuits, we 
need to add a clock, 
since the process of errors depends on time.
%
A quantum circuit will be associated a {\it timing}, 
 which indicates
the time step each  gate is applied.
The timing should be consistent with 
the circuit: if a gate gets as input output of another gate,
then it should be applied at least one time step after the other gate.
In a faultless computation, one can assume without loss
of generality that all the qubits are initialized and output together.
In the presence of faults, it is still not clear whether
allowing to  input constants  at any time 
helps the computation, since using qubits for a limited time might
protect them against faults.
Taking the stronger model, 
we permit also that qubits are input and output at different
times, and we say a qubit is {\it alive}
from $t_{1}$ to $t_{2}$ if it is input to the circuit 
at   $t_{1}$ and output at $t_{2}$.
We will assume all timings start at $0$, and denote by $T$
the last time step. 

We would like to incorporate in the model single qubit faults.
With out loss of generality, a single qubit fault,
 $F$, operating on the $q'th$ qubit
in the density matrix 
$\rho$ of $n$ qubits, can be described by adding a qubit to the system,
that represents the environment,
and letting a quantum gate to operate on it and the q'th qubit,
and then taking the reduced density matrix to the original qubits.
The fault is thus specified by the gate.
The resulting density matrix is denoted as $F^{q}\circ\rho$.
In this model,
a live qubit undergoes the fault $F$
with some probability $\eta$ per time step,
and $\eta$ is called the {\it decoherence rate}.
(One can allow
 the fault to depend 
on the time and on the exact qubit. 
All the results of this paper still hold.)

We now define the medium that realizes a quantum circuit 
with a certain timing, and is subjected to a certain
decoherence rate:
\begin{deff}
A quantum medium $M(Q,\tau,\eta\})$
 is a triplet: A quantum circuit $Q$,
 a timing $\tau$ on $Q$, and a decoherence rate $0\le \eta\le 1$.
\end{deff}


We define a fault path, which is a possible list of where and when 
the faults occur:
\begin{deff}
Let $M(Q,\tau,\eta)$ be a medium of $n$ qubits, 
which lives $T$ steps.
Denote by $t^{q}_{1}$,$t^{q}_{2}$ the times of
berth and death of the q'th qubit.
A {\it fault path} is a set of different pairs  $(q_{j},t_{j})$,
where  $1\le q_{j} \le n$,
 $t^{q_{j}}_{1}\le t_{j}\le t^{q_{j}}_{2}$.
\end{deff}

We associate with each path a weight, which is the probability 
for it to occur:
\begin{deff}
Let $M(Q,\tau,\eta)$ be a medium of $n$ qubits which has
$V$ pairs $(q,t)$ where faults can occur.
The weight of a path $\sigma$ is  
$w(\sigma)=d^{|\sigma|}(1-d)^{V-|\sigma|}$.
\end{deff}  

 $M$ computes a probabilistic function, where the distribution $f(i)$
which $M$ computes for an input $i$ is defined as a weighted average over
the outputs which the final density matrix of each fault path generates.
To formalize this, 
 we need notions of development in time of the medium.
For a medium of n qubits, an evolution is a possible such
  development, i.e.
 a list 
$\{E(i)\}_{i=0}^{T}$ of density matrices of $n$ qubits. 
There are special evolutions which develop
 subjected to a certain fault path,
and besides that ``obey'' the gates:

\begin{deff}
let $M$ be a medium of $n$ qubits,
$F$ a fault,
$\sigma$ a fault path.
An evolution $E$ is called
 a ``$\sigma,F$-trajectory of $M$'' 
if it satisfies $E(t+1)=
F^{\sigma_{t}}\circ \prod_{k}g^{k}_{t}\circ E(t)$
where $g^{k}_{t}$ are the gates applied 
 at time $t$,
and $F^{\sigma_{t}}$ are faults 
 applied in $\sigma$ at time t.
\end{deff}


We can now define the function computed by the medium 
when subjected to faults $F$:
\begin{deff}
Let the medium $M$ be subjected to faults $F$.
For an input string $i$ let
$\rho_{i}=\sum_{\sigma} w(\sigma)E_{i}^{\sigma}(T)$,
where $E_{i}^{\sigma}$ is an $F,\sigma$-trajectory of $M$
which satisfies $E_{j}^{\sigma}(0)=\rho_{|i\rangle}$.
The probability for $M$ to output a string $j$
for the input $i$ is the probability to get an outcome $j$ 
when applying basic  measurements on
 the result qubits in $\rho_{i}$.
\end{deff}

\subsection{\bf Equivalence to a weak interactions}
In the weak interactions model, each qubit undergoes a {\it weak fault},
deterministically, each time step.
The weak fault operates on a density matrix in the following way:
$WF^{q}\circ\rho=(1-\eta)\rho+\eta F^{q}\circ\rho$.
where $F$ is a single qubit fault, and $\eta$ is the strength of the
interaction with the environment.
The first term serves as a guard, so that the change in the density matrix 
is limited. 
To show that this model is equivalent to the probabilistic model,
it suffices to show that there is no difference between
 a mixed state which is a weighted sum of mixed states, and
 a mixed state prepared by choosing,
with the correct probability, one of the mixed states in the sum.
\begin{lemm}\label{trick}
Let $\rho=p_{1}\rho_{1}+p_{2}\rho_{2}$. 
The mixed state $\rho$ is quantumly equivalent 
to a mixed state 
prepared by choosing with probability $p_{i}$ the density matrix
$\rho_{i}$.
\end{lemm}
{\bf Proof:}
The distinguish ability of quantum states is by measurements.
By definition, the statistics  of any measurement will be the same
in the two cases. \bbox 
\subsection{\bf Equivalence to a model with many possible faults} 
We can also define a model which seems more physical:
At each time step, and for each live qubit, 
different possible faults $F_{i}$, can occur 
 with probabilities $\eta_{i}$, where $\eta=\sum_{i}\eta_{i}$.
It might seem surprising  at first sight, but
this model is equivalent to our probabilistic model
since we can define the one fault in our model to be a probabilistic 
combination of all the faults: 
 \(F\circ\rho=\frac{1}{\eta}\sum_{i}\eta_{i}F_{i}\circ\rho\).
\section{Polynomial simulations of sequential mediums}
We describe a simulation of a quantum compuation by a quantum medium, on
a randomized classical Turing machine.
As  described in the introduction, 
it suffices for our purposes to
regard only collapse faults.
We show that  the expected   
 slowdown of the simulation,
for sequential mediums and mediums which
operate gates only on $O(log(n))$ qubits each time step,
is {\em polynomial} in the time and number of qubits of the medium,
for any non-zero decoherence rate, $\eta>0$.

To study the effects of decoherence on QTM one must refer to some
possible physical realizations of QTMs.
 Our results apply immediately to the 
possible realization of QTM by sequential Quantum circuits,
 suggested by Yao\cite{Yao93}.  
 At the end of this section we informally describe how to extend the
results to general multi-head QTM.

\subsection{\bf A classical simulation of quantum mediums}\label{sim}
We define an algorithm $A(M,F)(i)$  which gets as an input
a description of a Quantum medium
  $M$, a single qubit observable $F$, and an input string $i$,
 and outputs the function that the
medium outputs when subjected to collapses of type $F$.  
$A$ presents a density matrix, $\rho$, of $n$ qubits,
  by a {\it configuration}: a list of density matrices,
$\rho_{j}$, each $\rho_{j}$  describing
 a set of entangled qubits, or a ``cluster'', 
 such that there is  no entanglement 
between different clusters.
 $\rho$ is the tensor product of the matrices 
in the configuration.
let us describe the probabilistic algorithm $A$:

\noindent
{\bf Input:}

 Description of a medium $M=(Q,\tau,d)$,

 string of $n$ qubits, $i$.

The eigenstates $|e_{1}\rangle, |e_{2}\rangle$,
of the observable $F$.

\noindent
{\bf Algorithm:}

- $A$ chooses a decoherence path $\sigma$ with probability
$w(\sigma)$, according to the decoherence rate $\eta$. 

- $A$ initializes the configurations $conf(0)$,$conf^{*}(0)$
 to be a list of matrices
  of one qubit, $|0\rangle\langle 0|$ or  $|1\rangle\langle 1|$
that correspond to $i$.

- For t=1 to T, $A$ does:

a) To simulate the computation step: 
Let $g$ be a gate that operates at time $t$.
Let $\rho^{q_{j}}$ be the matrices in $conf(t)$
that describe qubits on which $g$ operates. 
$A$ replaces them in the configuration
by the matrix  $g\circ\left(\rho^{q_{1}}\otimes...\rho^{q_{l}}\right)$.
For two qubits in the same cluster described by $\rho^{q}$,
we take $\rho^{q}$ in the tensor product only once. 
 This is done for all the gates 
of the $t'th$ time step
to give  $conf^{*}(t)$. 

b) To simulate a decoherence step: Let $(q,t)$ be a pair in $\sigma$.
Let $\rho_{q}$ be the matrix in $conf^{*}(t)$ that describes
the qubit $q$, and maybe a set $B$ of more qubits. 
 $A$ computes, from $\rho_{q}$, 
the probability for $q$ to collapse on each of the eigenvalues of $F$.
Than $A$  tosses a coin with this bias, to decide on which
of the eigenstates the qubit collapses.
Let the result be $|e_{1}\rangle$. 
$A$ replaces $\rho_{q}$ in $conf{*}(t)$
by the matrix $|e_{1}\rangle\langle e_{1}|$,
and the reduced  matrix $F_{e_{1}}\circ(\rho_{q})|_{B}$
This  simulates a  measurement of the $q'th$
qubit, {\em conditioned} that the outcome was $e_{1}$.

\noindent
{\bf Output:}
 To output the results of the computation,
$A$ imitates basic measurements of the result qubits,
as in the decoherence step, and outputs the outcome.

{~}

To show that $A$ computes the same function
as the quantum medium, we observe that
the simulation actually computes the $F,\sigma-trajectory$
of the medium, for the path it chose, except one thing:
The simulation makes a {\it conditioned} measurement,
and continues according to the result,
while the effect of the measurement faults is of an {\it unconditioned}
measurement. 
However:

\begin{lemm}
The output distribution  of $A(M,F)(i)$ equals the distribution
 $f_{M}(i)$ that $M$ computes when subjected to faults  $F$,
for an input $i$.
\end{lemm}
{\bf Proof:} By lemma \ref{trick},
the mixed state $F^{q}\circ\rho=P_{1}|e_{1}\rangle\langle e_{1}|\otimes
F_{e_{1}}\circ\rho|_{B}+P_{2}|e_{2}\rangle\langle e_{2}|\otimes
F_{e_{2}}\circ\rho|_{B}$
can not be 
distinguished from a mixed state prepared from 
$|e_{i}\rangle\langle e_{i}|\otimes F_{e_{i}}\circ\rho|_{B}$ with probability 
$p_{i}$. 
Therefore the effect of the unconditioned measurement
is the same as taking with the correct probabilities the
conditioned measurements.
\bbox

\subsection{\bf The simulation is efficient.}
To analyze the cost of the simulation,
 we divide the qubits each time step to two sets:
The {\it individual qubits}, that in the configuration of $A$ at time $t$,
are described by  their own density matrix,
and the other qubits, which are entangled between themselves.
Each individual qubit is described in the configuration by a
density matrix with $4$ entries, but the description of
the entangled qubits might be exponential in their number.
Denote 
by $K(t)$, $K^{*}(t)$ the sizes of sets of the non-individual qubits 
at time $t$ in {\it $conf(t),conf^{*}(t)$} in the simulation.
To bound the cost of the simulation from above, 
we first prove
that the probability for the number of non-individual qubits
 to be large is  exponentially small. 
\begin{lemm}\label{hasam_expTM}
Let $M$  be a medium which operates on no more than
 $P$ qubits each step,subjected to collapses of type $F$, with $\eta>0$.
 Consider $A(M,F)(i)$.
 There exist constants $a>4$, and $b>0$ such that
 $Pr(K(t)\ge bP+j)\leq \frac{1}{a^{j}}$,
$Pr(K^{*}(t)\ge bP+j+P)\leq \frac{1}{a^{j}}$.
\end{lemm}
{\bf Proof:} 
We use induction on $t$.
For $t=0$, $K(0)=0$, $K^{*}(0)\le P$.
For the induction step, assume $Pr(K(t)\ge bP+j)\leq \frac{1}{a^{j}}$ and
$Pr(K^{*}(t)\ge bP+j+P)\leq \frac{1}{a^{j}}$ for $t$.
To analyze $Pr(K(t+1))$, 
We apply decoherence on $K^{*}(t)$.
The probability to reduce the number of non-individual qubits
 is binomial so:
\[Pr(K(t+1)\ge bP+i)=\sum_{j=0}^{\infty}Pr(K(t+1)=bP+i+j)=\]
\[\leq \sum_{j,l=0}^{\infty}\frac{1}{a^{i+j+l-P}}
 \left( \begin{array}{c}bP+i+j+l\\l \end{array} \right) \eta^{l}(1-\eta)^{bP+i+j}=\]
\[= \sum_{j=0}^{\infty}\frac{(1-\eta)^{bP+i+j}}{a^{i+j-P}}
\left(\sum_{l=0}^{\infty}
 \left( \begin{array}{c}bP+i+j+l\\l \end{array} \right)
\left(\frac{\eta}{a}\right)^{l} \right)=\]
\[=\frac{a^{P+1}}{a-1}\left(\frac{(1-\eta)a}{a-\eta}\right)^{bP}
 \left(\frac{1-\eta}{a-\eta}\right)^i.\]

We want this to be smaller than $\frac{1}{a^{i}}$,
for any $i$. This implies two demands:
(1) \(\frac{1-\eta}{a-\eta}\leq \frac{1}{a}\), which
is achieved for $a\ge 1$. 
(2) The coefficient of the geometrical sequence
 must be smaller then $1$.
There exists $b$ that fixes that since 
 $\frac{(1-\eta)a}{a-\eta}$ is strictly smaller than $1$, because $\eta>0$.

For $Pr(K^{*}(t+1))$ we have that
$K^{*}(t+1)\le K(t+1)+P$. Then
\(Pr(K^{*}(t+1)\ge bP+i+P)\le \)
\(Pr(K(t+1)\ge bP+i)\leq \)\( \frac{1}{a^{i}}.\)\bbox

We can now prove that the expected number of entries of
density matrices that $A$ writes 
is polynomial:
\begin{lemm}\label{expec_exp}
Let $M(Q,\tau,\eta)$ be a medium of n qubits with parallelism  
$O(log(n))$, $\eta>0$,
$F$ a measurement fault, $i$ an input string.
 Consider $A(M,F)(i)$. The expected number of entries
 of the density matrices in all the configurations that  $A$ writes  
 is polynomial in n and $T$.
\end{lemm}

{\bf Proof:}
We use the definitions of the constants $a$ and $b$ 
 from lemma \ref{hasam_expTM}.
We compute the expectation value of $m(t)$, the number of entries for 
 the matrices of the $K(t)$ non-individual qubits,
at time $t$:
\(E(m(t))= \sum_{m=1}^{\infty} Pr(m(t)=m) m
\le\sum_{j=0}^{\infty} Pr(K(t)=j) 4^{j}\)
\(\leq\sum_{j=0}^{bP-1}   4^{j} +
\sum_{l=0}^{\infty} \frac{1}{a^{l}}  4^{bP+l}\), 
where the last inequality is by lemma \ref{hasam_expTM}.
The sum is finite since $a>4$.  The expectation is constant if
the parallelism,$P$, is constant, and
for $P=clog(n)$, 
the expectation is $\theta(n^{2cb})$. 
At time $t$  we also have to write $4$ entries for each small
density matrix, which adds at most $4n$ entries. 
The expected total number of entries, (during the whole simulation)
is the sum of the expectations in each time step. 
Thus it is polynomial in $n$ and linear in $T$. 
\bbox

\begin{theo}\label{TM}
Let $M$ be a medium of n qubits, life length $T$,
with $\eta>0$, and  $O(log(n))$ parallelism, $F$ a measurement fault.
 The expected cost of  $A(M,F)(i)$ will be polynomial in $T$ and $n$. 
\end{theo}
{\bf Proof:} By lemmas \ref{hasam_expTM},\ref{expec_exp} and the fact
that  one entry of a density matrix can be written in poly(n,T) bits,
  if the specification of the gates is
by poly(n,T) bits. \bbox

%


\subsection{\bf Simulating quantum Turing machines}
We briefly mention how to extend the above result to 
other possible models of QTM.
We assume a decoherence model for multi-head QTM
where besides single site faults, we allow faults on
the heads' positions  and the processor's state.
Again we refer to the model of random collapses, where we can show: 

\begin{theo}\label{TM2} 
Let $Q$ be a QTM, subjected to measurement faults
$F$ of the cites, the heads' position and the machine's state, with
 $\eta>0$.  A classical probabilistic Turing  machine can
simulate $T$ computation steps of $Q$ on inputs of
size $n$ in time polynomial in $T$ and $n$.
\end{theo} 

{\bf Proof:} Sketch. After every constant number of steps, the head's position
will collapse to one place. Since at each time step it can only triple the
number of sites it works on (by going Left, right or staying in place) the
expected number of different paths of the head  at each moment will remain
constant. The simulation will remain efficient
 when the number
of heads is $O(\log n)$ since then the number of paths is
polynomial.

\section{Simulating mediums with general parallelism}

In this section we study the effects of decoherence on
the computational power of general Quantum mediums, by
applying the simulation $A$, defined in 
section \ref{sim}, for mediums with general parallelism.
We assume gates of fan-in $\leq 2$, but
 the results hold for any constant fan-in. 
We show analytically that the cost of the simulation has
an expected polynomial slowdown for any medium with high enough decoherence.  
If the medium uses gates on more then one qubit not too frequently,
the simulation is still polynomial even for lower decoherence rates.
The simulation is proved to be exponential for a general random medium with
low enough decoherence rate. 
Finally we present computer experiments, showing a phase transition 
in the expected cost at some decoherence rate, where the 
cost ``jumps'' from polynomial  to exponential.
These transition points are found for random circuits and 
nearest neighbor circuits.
Known results~\cite{fri} imply that for random circuits the transition 
happens in a range of decoherence rates which is $O(\frac{1}{log(n)})$. 
For  any sequence of quantum circuits for which
there is a gap between the costs for high and low decoherences,
the transition happens in a range which goes to zero as the size
of the circuit goes to infinity.

%

\subsection{\bf Mediums with high decoherence rate.}

It is proved that for high enough decoherence, the expected 
slowdown of the simulation $A$ is polynomial.
We need a more refined analysis than in the sequential case, since 
the line of reasoning applied there 
leads in the general parallelism case to
a trivial exponential 
 upper bound on the cost.
Recall that a cluster in a configuration of $A$ is
a set of qubits described by the same density matrix in
 the configuration.
In the analysis of the sequential case we referred to the
total number of non-individual qubits, i.e qubits
which are in clusters of size bigger than one.
Here we take a closer look and consider the sizes of each cluster
alone.
We define $K_{q}(t)$,$K_{q}^{*}(t)$ as the size
of $q$'s cluster in
 $conf(t),conf^{*}(t)$, respectively.
As long as the sizes of the clusters are $O(log(n))$, the simulation is 
polynomial in $n$, since the number of entries in each density
matrix in $conf(t)$ will be polynomial.

We intend to show by induction on $t$ 
that the probability for $K_{q}(t)$ to be large 
 decays exponentially.
The difficulty here is to bound the change in the 
clusters' sizes during  the computation step.
If we could bound this, so that the clusters do not
grow too much during the computation step,
and if the decoherence is high enough to shrink them back down,
then the induction step will work.
In the following we  analyze separately the 
 change in the distribution of sizes of clusters,
during the computation step, and during
the decoherence step.
 Then we combine the two
to show that the cost of the simulation is polynomial 
for high enough decoherence. 

\subsubsection{Clusters' Growth in a computation step}
Assuming an exponential decaying bound on the sizes of clusters
before the computation step, we can give
 an  exponential decaying bound 
on the sizes of clusters  
after the computation step.

To do that we observe a connection to a branching process.
Let the clusters in the configuration at time $t$, $conf(t)$,
 be $A,B,C,..$.
Let $M$ be the next computational step.
The question is what is the cluster's size of
 a qubit $a\in A$ after $M$ is applied.
 Recall that we 
assume gates of fan-in two at the most. 
If a gate in $M$ operates on two qubits then they are called {\em mates}. 
We associate with $a$ a {\it tree} of which $a$ is the {\it root}.
$a$'s sons in the tree are the qubits in $A$.
Suppose $a_1$ is a son of $a$ and a mate of $b\in B$.
Then all members in $B$ are the sons of $a_1$,
and so on.
In each stage we ignore qubits that already appear
 in the tree. 
The distribution on sizes of clusters corresponds to 
a distribution on the number of sons in a branching process.
The distribution on the sizes of clusters after the computation
corresponds to the sizes of trees in the branching process.
If $a$ has no mate, $K_{a}^{*}(t)$, which is
 the size  of $a$'s cluster after $M$ is applied,
 equals the size of $a'$s tree.
If $a$ has a mate, $c$, then $K_{a}^{*}(t)$
 is the sum of sizes of $a$'s tree and of $c$'s tree,
where again duplicated qubits are ignored.

We first consider a simplified version of the 
above process where
(1) All gates in $M$ are of fan-in 2, and
(2) Gates in $M$ operate only on qubits from different clusters.
With these simplifications, the process that generates a tree,
defined above, turns out to be a simple   
 branching process, for which we can compute the 
probability distribution over the total size of two trees,
given the distribution on the number of sons.

\begin{lemm}\label{br}
Let $B$ be a branching process, with 
$Pr(z=j)\le \frac{1}{a^{j}},a>8$ for
 $z$, the number of sons.  
Let $T_{a},T_{b}$, be the sizes of two independent trees
 generated by $B$. Let  $L=2T_{a}+2T_{b}-2$. Then
 \(Pr(L=i)\leq a(\sqrt{2}-1)^{2}(\sqrt{\frac{8}{a}})^{i}.\)
\end{lemm}

{\em Proof:}
For the standard proofs in the following discussion, consult \cite{Alon}.
Let the generating functions\footnote{The generating function $G_{A}(x)$
of a random positive integer $A,$ 
is the expectation of $x^A$: $E(x^{A})=\sum_{i=0}^{\infty}Pr(A=i)x^i$}
 of $z,T,L$ be $G_{z},G_{T},G_{L}$.
In the following  formulas, the first\cite{Alon}
 connects the generating functions
 of the size of a tree $T$ to the number of sons $Z$ By definition of $L$  we have. The second  we have by definition of $L$:
\begin{equation}\label{LisTsq}\label{connect_zT}
G_{L+2}(x)=(G_{T}(x^{2}))^{2}~ ,~
G_{T}(x)=xG_{z}(G_{T}(x))
\end{equation}
We don't know the generating function for the variable $z$
since we only have a bound on the distribution.
Let us refer to the exponential decaying bound as 
a pseudo distribution, although it is not normalized to one,
and define ``pseudo generating functions'', (denoted by a tilde),
such that $\tilde{G}_{z}(x)$ will ``generate'' the exponential decay
 $\frac{1}{a^{i}}$: 
\(\tilde{G}_{z}(x)=\sum_{i=0}^{\infty}\frac{1}{a^{i}}x^{i}= \frac{a}{a-x},\)
and $\tilde{G}_{T}(x)$, $\tilde{G}_{L}(x)$ will satisfy equations
\ref{LisTsq}:
\(\tilde{G}_{T}(x)=x\tilde{G}_{z}(\tilde{G}_{T}(x))~,~
\tilde{G}_{L+2}(x)=(\tilde{G}_{T}(x^{2}))^{2}.\)
The coefficients $z_{i},t_{i},l_{i}$ in the analytic expansion
 of  $G_{Z}(x)$
$G_{T}(x)$, and $G_{L+2}(x)$ are smaller then the corresponding coefficients 
 $\tilde{z}_{i},\tilde{t}_{i},\tilde{l}_{i}$
of the ``pseudo generating function,'' when the expansions exist.
To prove this, note that
 the assumptions of the lemma imply $z_{i}\le \tilde{z}_{i}$.
 Writing equation \ref{connect_zT} using analytic expansions,
one can show by induction that $t_{i}\le \tilde{t}_{i}$.
Using this  and writing  equation \ref{LisTsq} in analytic expansion,
we have $l_{i}\le \tilde{l}_{i}$.

We intend to give an upper bound on $\tilde{l}_{i}$, 
since this will imply an upper bound
on $l_{i+2}=Pr(L=i)$ and therefore on $Pr(L\ge i)$.
Given $\tilde{G}_{z}(x)$, we can solve 
\(\tilde{G}_{T}(x)=x\tilde{G}_{z}(\tilde{G}_{T}(x))\)
 for $\tilde{G}_{T}(x)$, which
 gives a quadratic equation.
 We choose the ``minus'' 
solution of the quadratic equation, which corresponds to the 
correct choice in the case of $G_{T}(x)$.
It gives  \(\tilde{G}_{T}(\frac{a}{8})=\frac{(\sqrt{2}-1)a}{\sqrt{8}}\),
which implies
\begin{equation}
 \frac{(\sqrt{2}-1)^{2}a^{2}}{8}=(\tilde{G}_{T}(\frac{a}{8}))^{2}=
\tilde{G}_{L+2}(\sqrt{\frac{a}{8}}).
\end{equation}
 $\tilde{G}_{T}(x)$ is analytic for $|x|<\frac{a}{4}$,
so $\tilde{G}_{L+2}(y)$ is analytic at $y=\sqrt{\frac{a}{8}}$,
and we can write:
\(\frac{(\sqrt{2}-1)^{2}a^{2}}{8}=
\sum_{i=0}^{\infty} \tilde{l}_{i} (\sqrt{\frac{a}{8}})^{i}\),
which gives \(\tilde{l}_{i}\le (\sqrt{2}-1)^{2}\frac{a^{2}}{8}
\left(\sqrt{\frac{8}{a}}\right)^{i}\) that implies the desired result.
\bbox

We use the connection to the branching process defined above, to 
show an upper bound on the clusters' size after the computation step,
given the upper bound before the computation step.
\begin{lemm}\label{QC1}Let $a>8$, and  
\(C=(\sqrt{2}-1)^{2}\frac{a^{\frac{3}{2}}}{\sqrt{a}-\sqrt{8}}\).
 \(Pr(k_{q}(t)\ge j)\leq \frac{1}{a^{j-1}}\)  implies
\(Pr(K^{*}_{q}(t)\ge i)\leq 
C(\sqrt{\frac{8}{a}})^{i}.\)
\end{lemm} 
{\bf Proof}:  
Compare our process to the simplified branching process,
where identifying $k_{q}(t)-1$ with the variable $z$ and 
$k^{*}_{q}(t)$ with $L$.
Our process differs in two aspects:
It might connect mates inside a cluster,
and a qubit might not have a mate at all, i.e might not have a chance to
give birth to children.
The only effect of these fact can be to reduce the final cluster. \bbox

\subsection{\bf The decoherence step}

%
We now  show that with strong enough decoherence rate,
the distribution over sizes of clusters is  ``pulled'' by decoherence
back below the original exponential decaying bound:
\begin{lemm}\label{QC2}
 \(\exists ~\eta_{1}<1,~ a>8\) such that $\forall~ \eta>\eta_1$  
  \(Pr(K^{*}(t)\ge i)\leq C(\sqrt{\frac{8}{a}})^{i}\)
with
\(C=(\sqrt{2}-1)^{2}\frac{a^{\frac{3}{2}}}{\sqrt{a}-\sqrt{8}}\), implies
 \(Pr(k(t+1)\ge l)\leq \frac{1}{a^{l-1}}.\)

\end{lemm}
{\bf Proof:}
We find $\eta_1<0.97$.
We have to calculate how a distribution over sizes of clusters changes 
because of the decoherence.
The probability to reduce the number of qubits in a cluster is binomial
so:
\[Pr(k(t+1)\ge l)=\sum_{r=0}^{\infty}Pr(k(t+1)= l+r)=\]
\[\sum_{r=0}^{\infty}\sum_{p=0}^{\infty}Pr(K^{*}(t)=l+r+p)
 \left( \begin{array}{c}l+r+p\\p \end{array} \right) 
\eta^{p}(1-\eta)^{l+r}\leq\]
\[\sum_{p=0}^{\infty}Pr(K^{*}(t)\ge l+p)
\sum_{r=0}^{\infty} \left(\begin{array}{c}l+r+p\\p \end{array}\right) 
\eta^{p}(1-\eta)^{l+r}\leq\]
\[\le C(\sqrt{\frac{8}{a}})^{l}\sum_{r=0}^{\infty}(1-\eta)^{r+l}
\sum_{p}^{\infty}(\eta\sqrt{\frac{8}{a}})^{p}
\left(\begin{array}{c}l+r+p\\p \end{array}\right)=\]
\[=C
(\sqrt{\frac{8}{a}})^{l}\sum_{r=0}^{\infty}\frac{(1-\eta)^{r+l}}
{(1-\eta\sqrt{\frac{8}{a}})^{l+r+1}}=\]
\[=(\sqrt{2}-1)^2\frac{a^{\frac{5}{2}}}{\eta(\sqrt{a}-\sqrt{8})^2}
\left(\frac{\sqrt{8}(1-\eta)}{\sqrt{a}-\eta\sqrt{8}}\right)^l.\]


We want that for all $l$ $Pr(k(t+1)\ge l)\leq a(\frac{1}{a})^{l}$,
so we demand that the factor of the geometrical
sequence is smaller than $\frac{1}{a}$
and that the first term is  smaller then $a$.
The demands are satisfied with $a=30$ and $\eta\ge  0.97$.\bbox 
\subsubsection{Closing the  proof}
\begin{theo}\label{QC}
There exists a constant $\eta_1<1$ such that
for any medium $M$ on $n$ qubits, with decoherence rate $\eta>\eta_{1}$,
which lives for $T$ steps, and any measurement fault $F$ and input string $i$,
the expected cost of $A(M,F)(i)$
 is polynomial in $n$ and $T$.
\end{theo}

{\bf proof:}
 We show that for  $A$  of mediums  with
a decoherence rate  higher  than  $\eta_{1}$ 
defined  in lemma \ref{QC2},   and  for  any  qubit  q, 
the  sizes of $q$'s  clusters in  the algorithm  $A$ 
 satisfy 
for some
$a>8$ \(Pr(k_{q}(t)\ge j)\leq \frac{1}{a^{j-1}}\)  
and \(Pr(K^{*}_{q}(t)\ge i)\leq C(\sqrt{\frac{8}{a}})^{i}\)
for \(C=(\sqrt{2}-1)^{2}\frac{a^{\frac{3}{2}}}{\sqrt{a}-\sqrt{8}}\).
This is done by induction on $t$.
The basis of the induction is trivial, as
 $K_{q}(0)=1,   K^{*}_{q}(0)=1$ always.
The induction step is done in two stages,
by  lemmas  \ref{QC1} and  \ref{QC2}.
With   this exponential decaying bound  on  the distribution
over sizes of clusters, we  can proceed to show  
that  the  expected number  of  entries of  matrices
that  $A$ writes  is polynomial,  as  in  lemma \ref{expec_exp},
and  since   each entry  can be  written using a  polynomial
number   of bits, ( if  the gates are  specified in polynomial 
number,)  this completes   the  proof, with $\eta_{1}=0.97$.\bbox 
 
The numerical lower bound of $\eta_{1}=0.97$ is
 ofcource not a reasonable physical number,
to even start computation with;
It is high because it is good for any quantum  circuit.
Most circuits are simulated efficiently
with much lower decoherence rate(see section \ref{experiment}).

\subsection{\bf The importance of the frequency of many-qubits-gates} 
A quantum circuit that uses single qubit gates often,
is simulated efficiently already for low decoherence rates,
since single qubit gates do not participate in the struggle against
decoherence.
To see that, consider a quantum circuit that
applies a many-qubits gate on each qubit 
once every $\nu$ time steps at the most.
One can treat these $\nu$ time steps as a one time step,
in our branching process analysis.
During this ``extended'' time step,
the decoherence had it's chance to operate $\nu$ times.
The {\it effective} decoherence rate,
i.e. the rate of decoherence with respect to the extended
time step,
 is therefore $1-(1-\eta)^{\nu}\gg \eta$. 
Therefore already for very small decoherence rates $\eta$
the effective decoherence will be high enough for the
simulation to be efficient.

\subsection{\bf The simulation is exponential for
low decoherence in random circuits}
In this subsection we turn to the other side of the decoherence
rate scale, and ask what is the expected cost for low decoherence rates.
Off cource, there are circuits which can be simulated efficiently 
even with no decoherence at all. 
However, for random circuits the expected cost of the simulation is
exponential for $\eta<0.5$.  
By a  random circuit we mean a circuit in which  
 each time step
 a random matching between qubits is chosen,
and this matching corresponds to gates of fan-in two operating on the qubits.
\begin{lemm}
The simulation of a random circuit
subjected to decoherence rate $\eta<0.5$ is exponential.
\end{lemm}
{\bf Proof:}
We prove that when the decoherence rate is $\eta<0.5$
there is a cluster of size linear in $n$.
During one computation step
a cluster of $\alpha n$ qubits,
  will grow to
 $(\alpha+\alpha(1-\alpha))n$ since the probability for
 qubits in the cluster
to be matched outside the cluster is $(1-\alpha)$,
and these matching add at least one qubit to the cluster.
The decoherence step multiplies this by $\eta$.
Hence the original cluster size is multiplied each time
 step by a  factor $(2-\alpha)\eta$.
For $\eta> 0.5$ this is bigger than $1$ for small $\alpha$, and
the cluster will grow until it is of size   
$(2-\frac{1}{\eta})n$, which is linear in $n$.
The rate of growth of the cluster is exponential,
because 
as long as $\alpha$ is smaller than $2-\frac{1+\epsilon}{\eta}$,
the factor is bigger than $1+\epsilon$.

\subsection{\bf Phase transitions: experimental results}\label{experiment} 
We have  performed computer experiments to investigate 
how the  clusters sizes depend on the decoherence rate.
 To study the random algorithm case, we 
started with n clusters of size one, and each time step
chose a random matching between qubits.
A match represented a gate on two qubits,
so their clusters were joined.
To imitate faults, we randomly and independently separated a qubit from 
it's cluster with probability $\eta$, each time step.   
The system exhibits a {\it phase transition} at decoherence 
$\eta_0\approx 0.64$.
For $\eta>\eta_0$ the clusters' sizes stay smaller than $log(n)$,
but for  $\eta<\eta_0$ a  ``giant'' cluster of size
 linear in $n$ appears, which means that
 the density matrix describing it is exponential.

We also studied experimentally a one dimensional array 
of qubits, where gates where allowed to operate on nearest neighbors
only, and each qubit was matched with
it's left neighbor and right neighbor alternately. We found 
the transition point at  $\eta\approx 0.5$.

\section{Conclusions}
We feel that this work gives insight on the effects of 
decoherence on quantum computation,
and emphasizes the importance of $O(n)$ parallelism in a
 quantum computation
 subjected to decoherence processes.

Our results 
with respect to sequential quantum computers,
and quantum computers with $(log(n))$ parallelism
show that any amount of decoherence
destroys all the {\em quantum} computational power.

Considering parallel computers,
an upper bound was given on the decoherence rate which 
any quantum computer can hope to tolerate.
An important role  is given to the size of the fan-in used in the
computation, and to the frequency of many-qubits-gates.
Computers with many single qubit gates 
seem more sensitive to
decoherence, and can be simulated efficiently even
with very low decoherence rates. 
This should be considered when planning implementations of
quantum gates by low-fan-in gates\cite{DiV95,Bar95}.
It might also be interesting to quantify, according to these
observations, the stability of different algorithms against decoherence.

DiVincenzo\cite{DiV95} raises the question of whether the time evolution
of an open quantum system is a powerful computational tool.
Our conclusions are that 
at least within the model we have used, the answer is negative.

An open question is whether a stronger simulation than ours exists,
that is efficient for lower decoherence rates or even with no 
decoherence at all.

Early results\cite{Sho96,CS95} indicate
the possibility of an efficient simulation of an ideal QC by quantum mediums 
subjected to low decoherence rates.
Though our results might seem as a step towards negating the
fascinating possibility of a physical realization of
quantum computers, 
we actually view these results as optimistic, 
since phase transitions are known to be robust,
which implies that maybe quantum computers with low 
decoherence rates indeed can not be simulated efficiently.


\section{Acknowledgments}
We wish to thank Noam Nisan, Nati Linial, Yuval Peres, 
Ehud Friedgut, Amnon Ta-Shma, David DiVincenzo, Arthur Ekert, Avner Magen
 for very helpful conversations, essential remarks and corrections.
A special thank to Yoram Cohen for invaluable help in so many aspects.
This work was initiated during the Quantum Computation Workshop
at Torino, supported by the ISI Foundation.
\small
\bibliographystyle{plain}
\bibliography{references}
\end{document}